\documentclass[12pt]{article}

\usepackage{amsmath,amssymb}
\usepackage{graphics}

\newcommand{\bdm}[1]{{\mbox{\boldmath $#1$}}}

\numberwithin{equation}{section}
\newcommand{\be}{\begin{equation}}
\newcommand{\ee}{\end{equation}}
\newcommand{\bea}{\begin{eqnarray}}
\newcommand{\eea}{\end{eqnarray}}

\begin{document}
\baselineskip=16pt
\begin{titlepage}
\begin{flushright}
{\small OU-HET 596/2008}\\
%arXiv:0801.xxxx
\end{flushright}
\vspace*{1.2cm}

\begin{center}

{\Large\bf $\bdm{F}$-term Induced Flavor Mass Spectrum}\\

%{\Large\bf 
%Grand Unified Theory on Interval 
%} 
\lineskip .75em
\vskip 1.5cm

\normalsize
{\large Naoyuki Haba} 
 and 
{\large Yoshio Koide}$^\dagger$

\vspace{1cm}
{\it Department of Physics, Osaka University, 
1-1 Machikaneyama, Toyonaka, 560-0043, Japan} \\
{ [E-mail address: haba@phys.sci.osaka-u.ac.jp]} \\

\vspace{.3cm}
${}^\dagger${\it IHERP, Osaka University, 1-16 Machikaneyama, 
Toyonaka, 560-0043, Japan} \\
{ [E-mail address: koide@het.phys.sci.osaka-u.ac.jp]}

\vspace*{10mm}

{\bf Abstract}\\[5mm]
{\parbox{13cm}{\hspace{5mm}
%
%%%%%%%%%%%%%%%%%%%%%%%%%%%%%%%%%%%%%%%%%%%%%%%%%%%%%%%%%%%%%%%%
%             ABSTRACT                                         %
%%%%%%%%%%%%%%%%%%%%%%%%%%%%%%%%%%%%%%%%%%%%%%%%%%%%%%%%%%%%%%%%

New mechanism of generating flavor mass spectrum
 is proposed 
 by using an O'Raifeartaigh-type supersymmetry %(SUSY)
 breaking model. 
A desired bilinear form of fermion mass spectrum
 is naturally realized through $F$-components of 
 gauge-singlet (nonet of $SU(3)$ flavor symmetry) superfields, and 
 the suitable {charged-lepton} 
 mass relation is reproduced. 
The {\it charged-slepton}  mass spectrum 
 is non-degenerate in general, 
 and can be even hierarchical
 (proportional to the {charged-lepton}
 masses in the specific case).
Flavor changing neutral processes 
 are suppressed  
 since 
 the {charged-lepton} and {\it slepton} (except for 
 right-handed sneutrino) 
 mass matrices 
 are diagonalized simultaneously in the flavor space.
The right-handed {sneutrino}s are light with the similar 
 ratio to 
 the {lepton} sector
 ($\tilde{m}_{\nu_R}$/$\tilde{m}_e \sim m_\nu$/$m_e$).

%This model shows 
% the (next) lightest superparticle is 
% the (right-handed sneutrino) gravitino.  

%, which
% do not induce large flavor changing currents. 

}}

\end{center}

\end{titlepage}

%%%%%%%%%%%%%%%%%%%%%%%%%%%%%%%%%%%%%%%%%%%%%%%%%%%%%%%%%%%%
%                         Introduction                     %
%%%%%%%%%%%%%%%%%%%%%%%%%%%%%%%%%%%%%%%%%%%%%%%%%%%%%%%%%%%%
\section{Introduction}

Investigating an origin of a flavor mass spectrum will provide
 an important clue to the underlying 
 theory of quarks and leptons.
In the standard model, the flavor mass spectrum is
 originated in 
 the structure of Yukawa coupling constants $Y^f_{ij}$ 
 ($f=u,d,\nu,e$; $i,j=1,2,3$), where fermion 
 mass matrices $M^f$ are 
 given by $M_{ij}^f = Y^f_{ij} \langle H \rangle$
 ($H$ is the Higgs doublet). 
%: a vacuum
% expectation value (VEV) of the Higgs doublet, $H$).
%, with $v_H=\langle H^0\rangle$).
When there is a flavor symmetry in the mass matrices $M^f$,
 we have predictions for the masses and
 mixings.

On the other hand,
 there is another idea for the origin of the flavor
 mass spectrum, which is 
 originated in a structure of vacuum expectation values (VEVs)
 of scalar fields 
 (all couplings are of ${\cal O}(1)$). 
This paper would like to take this approach, and 
 one of the authors (YK) 
 has 
 proposed a prototype of such a model\cite{Koide90},
 where there are three Higgs doublets %scalars $\phi_i$ 
 with VEVs of        
 $v_i=\langle H^0_i\rangle$ 
 satisfying 
\be
v_1^2+v_2^2+v_3^2 =\frac{2}{3}(v_1+v_2+v_3)^2 .
\label{(1.1)}
\ee
It %The relation (1.1) 
 can lead to
 the charged-lepton mass relation\cite{Koidemass},
\be
m_e+m_\mu+m_\tau=\frac{2}{3}(\sqrt{m_e}+\sqrt{m_\mu}+\sqrt{m_\tau})^2 ,
\label{(1.2)}
\ee
where $M^e={\rm diag.}(m_e,m_\mu,m_\tau) \propto v_i^2$,  
 which gave a remarkable prediction of 
 $m_\tau = 1776.97$ MeV from 
 the observed values of $m_e$ and $m_\mu$.\footnote{
The observed value is 
 $m_\tau^{obs}=1776.99^{+0.29}_{-0.26}$ MeV\cite{PDG06}.
}
However, a model with multi-Higgs doublets causes the 
 large flavor changing neutral current (FCNC) 
 in general\cite{FCNC}. 
Therefore, we must change the scenario 
 in which the origin of the mass spectrum is
 separated from the electro-weak symmetry breaking. 
%It is nice if the origin of
% mass is coming from the VEV ($v_H$) of the conventional Higgs doublet, 
% while the origin of 
A typical example is the Froggatt-Nielsen (FN) model\cite{Froggatt},
 which has new scalars $\phi$  
 and their VEVs ($v_i$) induce 
 mass spectrum 
 which has nothing to do with the
 electro-weak symmetry breaking. 
For example, in Ref.\cite{Koide0705}, a $U(3)_F$-flavor symmetry
 has been introduced, where 
 $\phi$ is 
 regarded as a $U(3)_F$-nonet field. 
Then Yukawa interactions of the
 charged-lepton sector are induced from 
  a higher dimensional operator 
$\left[ y_0({\phi_{ik}\phi_{kj}}/{\Lambda^2}) L_j H_d E_i\right]_F$
 in the supersymmetry (SUSY) theory, where 
$\Lambda$ is the cutoff scale of
 this effective Lagrangian.\footnote{
%$\Lambda$ can be regarded as a composite scale as
% in Ref.\cite{}. 
Reference \cite{Ma-PLB07} proposed 
 a similar model but without higher 
 dimensional operators.}   
$L$ and $E$ are $SU(2)_L$-doublet and singlet lepton superfields, 
 respectively, and 
 $H_d$ is the Higgs doublet 
 which couples
 to the charged-lepton superfields. 
This model can avoid the FCNC problem.
However, 
% inputting only 
% bilinear term of FN field ($\phi_{ik}\phi_{kj}$) seems artificial,  
% and
 we cannot explain why
 only the bilinear term of FN field ($\phi_{ik}\phi_{kj}$) 
 is introduced in
 the superpotential.\footnote{
%It might be
% justified if we impose some flavor symmetries. 
%However, a
A continuous flavor symmetry induces 
 massless NG bosons, 
 while a discrete flavor symmetry cannot forbid 
 higher order terms,
 such as $\phi^4$ in $Z_2$-symmetry.
}

In this paper, 
 we propose another mechanism 
 which naturally induces bilinear form
 of mass spectrum. 
For example, %This model induces 
 the charged-lepton masses are induced through
 $F$-components of the
 superfield $\phi$ as 
\be
\left[ y_0\frac{\phi_{ij}^\dagger }{\Lambda^2} L_j H_d E_i\right]_D .
\label{(1.3)}
\ee
%The SUSY is broken by 
% $F^\dagger_\phi = -\partial W/\partial \phi \neq 0$.  
Since a $F$-component has mass dimension two,
 so that Eq.(\ref{(1.3)}) can be a good candidate 
 to reproduce the mass relation of Eq.(\ref{(1.2)}).\footnote{
A similar type of Eq.(\ref{(1.3)}),
 $[ y_0'\frac{\phi_{ij}^\dagger }
 {\Lambda^2} L_j H_u N_i]_D$,
 was
 used for tiny  
 Dirac neutrino masses
 in Ref.\cite{Murayama}, where  
%$L$, $E$, and 
$N$ is the right-handed neutrino and 
$H_u$ is the Higgs doublet 
 which constructs up-type Yukawa
 interactions. 
}
%We will try to build a model which can 
% naturally induce the 
% mass formula of Eq.(1.2)
The vacuum of this model induces 
 the mass relation Eq.(\ref{(1.2)}) as well as 
 SUSY-breaking, so 
 we can expect 
 fruitful byproducts, such as sfermion 
 mass spectrum.

%%%%%%%%%%%%%%%%%%%%%%%%%%%%%%%%%%%%%%%%%%%%%%%%%%%%%%%%%%%
%%%%%%%%%%%%%%%%%
\section{A Model}

We mainly focus on the lepton sector. 
An application to the quark sector might be possible. 
Let us start showing our model.

\subsection{Lagrangian}

We adopt the following 
 O'Raifeartaigh-type superpotential\cite{ORft}, 
\be
W(\Phi, A, B)= \sum_{f=u,d} \left( W_\Phi(\Phi_f) + 
\lambda_A^f {\rm Tr}[A_f\Phi_f\Phi_f] 
+\lambda_B^f {\rm Tr}[B_f\Phi_f\Phi_f] -\mu_f^2{\rm Tr}[\xi_f A_f]
\right) .  
\label{(2.1)}
\ee
The K\"{a}hler potential is set as  
\bea
{\cal K}& = & {\cal K}_0+{\cal K}_1+{\cal K}_Y, 
\label{(2.2)} \\
{\cal K}_0 &=& \sum_{f=u,d} 
 \left({\rm Tr}[A_f^\dagger A_f]+{\rm Tr}[B_f^\dagger B_f]
+{\rm Tr}[\Phi_f^\dagger \Phi_f] + H_f^\dagger H_f\right) 
+ L^\dagger L+E^\dagger E +N^\dagger N ,
\label{(2.3)} \\
{\cal K}_1 &=& -\frac{1}{\Lambda^2} \sum_{f=u,d}
 \left({\rm Tr}[A_f^\dagger A_f]^2
+{\rm Tr}[B_f^\dagger B_f]^2 \right) ,
\label{(2.4)} \\
{\cal K}_Y &=&
\frac{1}{\Lambda^2}
 {\rm Tr}\left[\left( y_A^d A_d^\dagger +y_B^d B_d^\dagger
\right) LH_dE\right] 
+\frac{1}{\Lambda^2}
 {\rm Tr}\left[\left( y_A^u A_u^\dagger +y_B^u B_u^\dagger
\right) LH_u N\right] +{\rm h.c.} ,
\label{(2.5)}
\eea
where 
$\Phi_f$, $A_f$ and $B_f$ are $U(3)_F$-nonet superfields.
$\xi_f$ is a $3\times 3$ numerical matrix while 
 all other couplings ($y$'s, $\lambda$'s) 
 are complex numbers (not matrices). 
$W_\Phi$ is a superpotential which contains  only 
 $\Phi_f$.
%In addition to the canonical 
% K\"{a}hler potential ${\cal K}_0$, 
% we introduce 
${\cal K}_1$ is introduced to realize 
 $\langle A_f\rangle =\langle B_f\rangle =0$ and %to make the scalars
 also avoids massless scalers as will be shown later.\footnote{
The minus sign of ${\cal K}_1$ is assumed to
 be derived by the underlying theory.
} 
% $A_f$ and $B_f$ massive as we see later.
${\rm Tr}[A_d^\dagger LH_dE]$ and 
 ${\rm Tr}[B_d^\dagger LH_dE]$ in 
 ${\cal K}_Y$ play crucial roles of 
 generating effective
 Yukawa interactions for the charged-leptons and neutrinos.

We take $R$-charge assignments as
 $R(A_f)=R(B_f)=2$ and $R(\Phi_f)=0$ in order to 
 forbid phenomenologically unwanted interactions, 
 ${\rm Tr}[A_f^2]$, 
 ${\rm Tr}[A_f^3]$, 
 ${\rm Tr}[A_f^2 \Phi_f]$, 
 ${\rm Tr}[A_d LH_dE]$, ${\rm Tr}[B_d LH_dE]$, and
 so on. 
We also introduce an additional 
 $Z_2$-symmetry, 
 under which only $\Phi_f$
 transforms as 
 odd-parity fields. 
It forbids the terms,
 ${\rm Tr}[A_f \Phi_f]$, ${\rm Tr}[B_f \Phi_f]$,
 ${\rm Tr}[\Phi_d LH_dE]$, and so on. 
As for a tadpole term,  
 $\mu'^2{\rm Tr}[\xi_f B_f]$ can be eliminated by
 the field redefinition of $A_f$ and $B_f$.\footnote{
Precisely speaking, we should determine the 
 K\"{a}hler potential 
  Eq.(\ref{(2.2)}) after
 this redefinition.                                                            
} 
%The formulation in this section is the same both in the charged
% lepton and neutrino sectors, so that 
We will omit the indices 
 ``$u$" and ``$d$"
 when they are obvious.
%we pay attention to the charged lepton sector, 

%Especially, we pay attention to the charged lepton sector hereafter.

\subsection{Vacuum of the model}

The scalar potential ($V_{}=-{\cal L}_{\rm scalar}$) is
 given by 
%For the specific case with (2.1) - (2.5), we obtain a simple expression
\be
{V}_{} = - \left(
{\rm Tr}\left[ F_A \frac{\partial W}{\partial A} \right]
+ {\rm Tr}\left[ F_B \frac{\partial W}{\partial B} \right]
+ {\rm Tr}\left[ F_\Phi \frac{\partial W}{\partial \Phi} \right]
 \right) .
\label{(2.6)}
\ee
$F$'s are calculated from the equations of motions  
 ($\partial {\cal L}/\partial F=0$) as 
\bea
&& F_\Phi^\dagger +\frac{\partial W}{\partial \Phi}=0,
\label{(2.7)} \\
&& F_A^\dagger + \frac{\partial W}{\partial A} 
-\frac{2}{\Lambda^2} \left( {\rm Tr}[A^\dagger A] F_A^\dagger 
+{\rm Tr}[F_A^\dagger A] A^\dagger \right)
+\frac{1}{\Lambda^2} y_A^* {\cal X}^\dagger =0 ,
\label{(2.8)} \\
&& F_B^\dagger + \frac{\partial W}{\partial B} 
-\frac{2}{\Lambda^2} \left( {\rm Tr}[B^\dagger B] F_B^\dagger 
+{\rm Tr}[F_B^\dagger B] B^\dagger \right)
+\frac{1}{\Lambda^2} y_B^* {\cal X}^\dagger =0 ,
\label{(2.9)} 
\eea
where ${\cal X}=F_L HE +LF_H E +L H F_E$. 
Notice that 
 $V_{}$ in Eq.(\ref{(2.6)})
 contains fields $L$, $E$ and $H$ through
 the K\"{a}hler potential, 
 ${\cal K}_Y$. 
%Their contributions are equivalent to 
% input 
% $%{\cal L} \ni 
% F_L^\dagger F_L +F_H^\dagger F_H + F_E^\dagger F_E$ 
% in the canonical K\"{a}hler potential  
% and eliminated by 
The equation of motion of $F_L^\dagger$
 is given by 
\be
F_L^\dagger +  \frac{\partial W}{\partial L} + 
\frac{1}{\Lambda}  HE(y_A F_A^\dagger + y_B F_B^\dagger) = 0 , 
\label{(2.10)}
\ee
and equations of motions of $F_H^\dagger$ and $F_E^\dagger$ are
 similar to it.\footnote{  %Eq.(\ref{(2.10)}).
Of cause Eqs.(\ref{(2.8)})$\sim$(\ref{(2.10)})
 are also obtained directly from the inverse of the 
 K\"{a}hler potential, such as   
%For example, 
%Eq.(\ref{(2.8)}) 
% is given by 
 $F_A^\dagger = -{\cal K}^{-1}_{AA^\dagger}dW/dA$.
}

It is hard to obtain exact solutions 
 of Eqs.(\ref{(2.8)}) and (\ref{(2.9)}), 
 thus we use a perturbation method 
 of expanding $1/\Lambda^n$. 
Then we obtain %This induces                                     
\be
F_A^\dagger = -\frac{\partial W}{\partial A} 
+\frac{2}{\Lambda^2} \left( {\rm Tr}
[A^\dagger A]\frac{\partial W}{\partial A}
+ {\rm Tr}\left[A \frac{\partial W}{\partial A}\right] A^\dagger \right)
+ {\cal O} \left(\frac{1}{\Lambda^4}\right) ,
\label{(2.11)}
\ee
%$$
%F_B^\dagger = -\frac{\partial W}{\partial B} 
%-\frac{2}{\Lambda} \left( {\rm Tr}[B^\dagger B]\frac{\partial W}{\partial B}
%+ {\rm Tr}\left[B \frac{\partial W}{\partial B}\right] B^\dagger \right)
%+ {\cal O} \left(\frac{1}{M^4}\right) ,
%\eqno(2.14)
%$$
and the similar equation of $B$ (which is
 calculated by a replacement of $A\rightarrow B$). 
%Now let us estimate the scalar potential $V$
% by using this perturbation method. 
As Eq.(\ref{(2.1)})
 derives 
\bea
\frac{\partial W}{\partial \Phi}& =& \frac{\partial W_\Phi}{\partial \Phi}
+\lambda_A (A\Phi+\Phi A) +\lambda_B (B\Phi+\Phi B),
\label{(2.12)}\\
\frac{\partial W}{\partial A}& =& \lambda_A \Phi\Phi - \mu^2 \xi,
\label{(2.13)}\\
\frac{\partial W}{\partial B}& =& \lambda_B \Phi\Phi ,
\label{(2.14)}
\eea
 the scalar potential
 becomes 
\bea
V& =&V_0 +V_1 ,
\label{(2.15)}\\
V_0 & =&{\rm Tr}\left[ \left| \frac{\partial W_\Phi}{\partial \Phi} +
\lambda (C\Phi +\Phi C)\right|^2 \right]
+ {\rm Tr}\left[ |\lambda_A \Phi\Phi - \mu^2 \xi|^2 
+ |\lambda_B \Phi\Phi|^2 \right] ,
\label{(2.16)}\\
V_1 & =& \frac{2}{\Lambda^2} \left\{ 
\left( |\lambda_A|^2 {\rm Tr}[A^\dagger A] + |\lambda_B|^2
 {\rm Tr}[B^\dagger B] \right)
{\rm Tr}[\Phi\Phi\Phi^\dagger \Phi^\dagger] \right. 
\nonumber \\
& & \left. -\left( \lambda_A^* \mu^2 {\rm Tr}[A^\dagger A]
 {\rm Tr}[\xi \Phi^\dagger \Phi^\dagger]
+{\rm h.c.} \right)
 +|\mu^2|^2  {\rm Tr}[A^\dagger A] %+{\rm Tr}[B^\dagger B])
 {\rm Tr}[\xi^\dagger \xi] \right.  \nonumber \\
 & & \left. + \left| \lambda_A {\rm Tr}[A\Phi\Phi] -\mu^2 {\rm Tr}[A\xi] 
\right|^2 
+ |\lambda_B |^2  | {\rm Tr}[B\Phi\Phi] |^2
\right\} , %\nonumber
\label{(2.17)}
\eea
where 
 $C$ and $C'$ are defined by
\be
\lambda C \equiv \lambda_A A +\lambda_B B,
\ \ \ 
\lambda C' = -\lambda_B A +\lambda_A B 
\label{(2.18)}
\ee
with $\lambda=\sqrt{\lambda_A^2+\lambda_B^2}$. 
Notice that 
 $V_0$ ($V_1$) is the potential of  ${\cal O}(\Lambda^0)$
 (${\cal O}(\Lambda^{-2})$).  

Nine scalars of $C'$ are massless 
 in the tree level potential $V_0$,  
 because 
 Eq.(\ref{(2.16)}) does not contain
 $C'$. 
Thus, in the direction of $C'$, 
 $\langle A \rangle$ and $\langle B \rangle$  
 are not determined at ${\cal O}(\Lambda^0)$. 
%The location of vacuum is not 
% determined at  ${\cal O}(M^{0})$ . 
They are 
 determined by  
 including 
 ${\cal O}(\Lambda^{-2})$ 
 corrections of $V_1$. 
Stationary conditions $dV/dA=dV/dB=0$
 decide the vacuum at 
\be
\langle A\rangle = \langle B\rangle = 0   
\label{(2.19)}
\ee
with the condition 
\be
 \frac{\partial W_\Phi}{\partial \Phi} = 0.            
%+(\lambda_A A+\lambda_B B)\Phi + \Phi (\lambda_A A+ \lambda_B B)=0 ,
\label{(2.20)}
\ee
Under the Eqs.(\ref{(2.19)}) and (\ref{(2.20)}), 
 the condition of 
  $dV/d\Phi =0$ determines 
 $\langle\Phi\rangle$ as 
\be
\langle \Phi\rangle\langle\Phi\rangle = 
 \frac{\lambda_A}{\lambda_A^2+\lambda_B^2} \mu^2 \xi .
\label{(2.21)}
\ee
Then the height of 
 the scalar potential 
 at this vacuum 
 is given by  
\be
V_{\rm min}=
%{\rm Tr}\left[ \left( \lambda_A 
%\frac{\lambda_A}{\lambda_A^2+\lambda_B^2} \mu^2 \xi
% -\mu^2 \xi\right)^2 \right] 
%+ \lambda_B^2 \left(\frac{\lambda_A}{\lambda_A^2+\lambda_B^2}
% \mu^2\right)^2 {\rm Tr}[\xi\xi] = 
\left( 1-\frac{\lambda_A^2}{\lambda_A^2+\lambda_B^2} 
\right) \mu^4 \; {\rm Tr}[\xi^\dagger\xi] .
\label{(2.18)}
\ee

We will show that the minimum exists at 
 $\langle L \rangle = \langle E \rangle =0$ 
 in Section 5, 
 and then 
 $F_\Phi^\dagger$, $F_A^\dagger$ and 
 $F_B^\dagger$ are given by
\be
F_\Phi^\dagger=-\frac{\partial W}{\partial \Phi} =0 , \ \ \ 
F_A^\dagger=-\frac{\partial W}{\partial A} = \lambda_B 
\frac{\lambda_B}{\lambda_A} \Phi\Phi , \ \ \ 
F_B^\dagger=-\frac{\partial W}{\partial B} = -\lambda_B \Phi\Phi,
\label{(2.22)}
\ee
{}from Eqs.(\ref{(2.7)})$\sim$(\ref{(2.9)}), respectively.
Thus, the effective Yukawa interaction of 
 the charged-lepton sector is 
 given by 
\be
 (y_A^d \lambda_B^d-y_B^d\lambda_A^d)\frac{\lambda_B^d}{\lambda_A^d} 
\frac{\langle\Phi_{dik}\rangle\langle
\Phi_{dkj}\rangle}{\Lambda^2} 
L_j H_d E_i  ,
\label{(2.23)}
\ee
which is the desirable
 bilinear form and might produce  
 the mass relation $m_{ei} \propto v_{di}^2$ 
 ($v_{di} = \langle \Phi_{dii}\rangle$)
 in the diagonal basis of
 $\langle\Phi_d\rangle$.

Notice that 
 Eq.(\ref{(2.21)}) seems to imply  
 the flavor indices of $\langle\Phi\rangle^2$ 
 are completely fixed by the parameter $\xi$. 
%This means that 
% the effective Yukawa coupling constant $Y_e$ defined by
% ${\rm Tr}[Y_e LH_dE]$ is fixed 
% when we give the parameter $\xi_d$. 
However, it is not correct because 
 Eq.(\ref{(2.20)}) must be also satisfied at
 the vacuum. 
We take a standpoint that 
 the parameter $\xi$ is basically free, and it is constrained
 by $W_\Phi(\Phi)$ through the relation 
 of Eq.(\ref{(2.21a)}) which will be 
 shown in Section 3. 
%We will investigate the form of 
% $W_\Phi(\Phi)$ in order to know the mass spectrum.

%%%%%%%%%%%%%%%%%
\subsection{Mass spectra of $\bdm{U(3)_F}$-nonet fields}

Now let us calculate fermion masses 
 of $U(3)_F$-nonet %(gauge singlet)
 particles 
 of $\Phi$, $A$ and $B$ 
 which are 
denote as $\psi_\Phi$, $\psi_A$ and $\psi_B$,
 respectively.
%In this section, we again drop the indecices $f=u,d$ except for 
% specific cases.
%,
% because the formulation is
% the same between those in the charged lepton and neutrino sectors.
The fermion masses are induced from
 the 2nd and 3rd terms
 in Eq.(\ref{(2.1)}) as 
\be
{\cal L}_{\rm mass} = \lambda \sum_{i,j} 
(\psi_{C})_{ij} (v_i+v_j) (\psi_\Phi)_{ji} 
+2 \lambda_d \sum_i (\psi_\Phi)_{ii} \sum_j \hat{v}_j (\psi_\Phi)_{jj}  ,
\label{(4.1)}
\ee
where $\hat{v}_i = v_i -(v_1+v_2+v_3)/3$ and 
$v_i=\langle\Phi_{ii}\rangle$. 
The superfields $A$ and $B$
 couple to $\Phi$ only through
 the term 
 ${\rm Tr}[(\lambda_A A +\lambda_B B)\Phi\Phi]$ 
 in the superpotential $W$ %of Eq.(2.1),
 so that  $\psi_{C'}$ are massless while $\psi_C$ 
 have Dirac masses of ${\cal O}(\langle\Phi\rangle)$
 with $\psi_\Phi$. 
This situation is not affected by the existence of the 
 K\"{a}hler potential ${\cal K}_1$ %of ${\cal O}(M^{-2})$   
 due to $\langle A\rangle =\langle B\rangle=0$.

%Are 
% these massless fermions $\psi_{C'}$ with
% flavor indices phenomenologically 
% acceptable?
By taking account of $R$-parity conservation,
 $A$ and $B$ should be regarded as $R$-parity even.
Thus the massless $\psi_{C'}$ are 
  the lightest superparticle (LSP). 
We should remind that 
 one degree of freedom 
 in   $\psi_{C'}$ is absorbed into
 the longitudinal mode of 
 the gravitino, and its mass becomes %whose mass is given by
 $m_{3/2} \simeq F/M_P$ ($M_P$: four-dimensional 
 Planck scale). 
Other eight components of 
  $\psi_{C'}$ are remaining as massless
 fermions,  
 which seem problematic from 
 the view point of phenomenology. 
But, is it true? 
$\psi_{C'}$ interacts with the 
 leptons through 
 the 1st and 2nd terms of 
 Eq.(\ref{(2.5)}) which contains 
 the interactions 
\be
{\cal L} \ni 
\frac{1}{\Lambda^2}({\rm Tr}[\partial_\mu \gamma^\mu \psi_{C'_d}^{\dagger}
\psi_L H_d E] + 
{\rm Tr}[\partial_\mu \gamma^\mu \psi_{C'_d}^{\dagger}
L \psi_{H_d} E]
+
{\rm Tr}[\partial_\mu \gamma^\mu \psi_{C'_d}^{\dagger} 
LH_d \psi_E]). %+\cdots .
\label{(4.3)}
\ee
%They seem to be dangerous since
% the next lightest particle (NLSP),
% maybe neutralinos $\tilde{\chi}^0$,
% decays to $\psi_{C_d'}$ through this
% interaction.
These interactions 
 in the diagrams with 
 $\psi_{C'}$ in the external lines 
 vanish by using the equation of motion,
 $\partial_\mu \gamma^\mu \psi_{C'_d} =0$.\footnote{
The interactions      
 $\frac{1}{\Lambda^2}({\rm Tr}[\psi_{C'_d}^{\dagger}
 \partial_\mu \gamma^\mu\psi_L H_d E] + 
{\rm Tr}[\psi_{C'_d}^{\dagger} L
 \partial_\mu \gamma^\mu\psi_H  E] + 
{\rm Tr}[\psi_{C'_d}^{\dagger} L H_d
 \partial_\mu \gamma^\mu\psi_E])$ 
 also vanish by using partial integral.} 
%Thus, the 
% NLSP ($\tilde{\chi}^0$) can be still a
% dark matter in our model.\footnote{
%If there is an interaction,  
% $\left[\frac{1}{M} {\rm Tr}[C'_d LH_dE] \right]_F$,
% a rapid decay  
% of the lightest neutralino is occurred, 
% for example through a process  
% $\tilde{\chi}^0 \rightarrow {\psi_L} +L \rightarrow 
% \psi_L+(\psi_{C'_d} +\psi_E)$,  
% however this term is forbidden 
% by the $R$-charge assignment.
%The decay process finishes before 
% 1 sec when $\Lambda \sim 10^5$ GeV.
%So the big bang necleon synthesis
% is not broken but another dark matter candidate
% is needed. 
%}
Thus, although $\psi_{C'}$ are massless LSP, 
 the decay processes to  $\psi_{C'}$ are 
 strongly suppressed. 
% the next-LSP (NLSP) might be stable.  
% as the dark matter. 
The FCNC processes, such as %$\mu\rightarrow 3e$,
 $\mu\rightarrow e\gamma$, 
 mediated by $\psi_{C'}$ in the loop diagrams 
 are also suppressed due to 
 the small coupling of (lepton)-(slepton)-($\psi_{C'}$) 
 and no chiral-flip of $\psi_{C'}$. %\footnote{
%Section 4 shows $\lambda={\cal O}(10^5)$ GeV 
% in which accurate estimations 
% are needed for the FCNC processes.
% %explicitly induces the enough suppression. 
%}    

% can exist in the one-loop diagrams,
% however the coupling 
% are enough suppressed 
% ($M^e\langle H_d \rangle/\Lambda^2$)
% so that FCNC processes mediated by $\psi_{C'}$
% are negligibly small.

Next let us estimate 
 the scalar masses of nonet fields. 
%As for 
% the scalar masses of $\Phi$, $A$ and $B$,
The scalar masses of $\Phi$       
  are of order $\langle \Phi \rangle$ which are 
 induced from 
\be
4\lambda_d^2 \left(\sum_i v_i^2\right) \left( \sum_j \Phi_{jj}\right)^2
+ 2\lambda^2 \sum_{i,j} (v_i v_j +\delta_{ij})
\Phi_{ij}\Phi_{ji} ,
\label{(4.5)}
\ee
in the diagonal basis of $\langle \Phi \rangle$.
On the other hand, 
 the masses of $C$ and $C'$ are of orders $\langle\Phi\rangle$
 and $\langle\Phi\rangle^2/\Lambda$, respectively,       
 which are induced  from 
\be
2\lambda^2 \left\{
{\rm Tr}[\langle\Phi\rangle \langle\Phi\rangle C C
+\langle\Phi\rangle C \langle\Phi\rangle C]
+\frac{\lambda_B^2}{\lambda_A^2} 
\frac{{\rm Tr}[\langle\Phi\rangle^4]}{\Lambda^2} 
\left({\rm Tr}[A A^\dagger]+({\rm Tr}[B B^\dagger]\right)
\right\} .   
\label{(4.4)}
\ee
It should be emphasized that 
% The important point is %We should notice 
% that %scalar masses of 
 $C'$ scalars become massive through %by the higher
% dimensional terms in
 $V_1$. 
Although $C$ and $C'$ interact with sleptons 
 through the K\"{a}hler interactions 
 $\frac{1}{\Lambda^2} {\rm Tr}[\partial^2{C}^{\dagger} {L} H_d {E}]$
 and 
 $\frac{1}{\Lambda^2} {\rm Tr}[\partial^2{C'}^{\dagger} {L} H_d {E}]$,
 respectively,  
 their contributions 
 to the FCNC processes are negligibly small.     
It is because these interactions only exit in
 higher loops as well as  
 $C$ and $C'$ are heavy enough.  
 
%Thus the decay widths of $C$ and 
% $C'$ to the leptons 
% are estimated as 
% $\Gamma \sim \langle \Phi \rangle^2\langle H_d \rangle/\Lambda^2$  
% and 
% $\Gamma \sim \langle \Phi \rangle^4\langle H_d \rangle/\Lambda^4$  

%has the decay width of 
% $\Gamma \sim \langle \Phi \rangle^4\langle H_d \rangle/\Lambda^4$  
% through 

% is harmless since
% the scalar components of $C'$ are heavy as shown below.

%The decay is faster than 
% 1 sec when $\Lambda \sim 10^5$ GeV. 

%%%%%%%%%%%%%%%%%%%%%%%%%%%%%%%%%%%%%%%%%%%%%%%%%%%%%%%%%
%%%%%%%%%%%%%%%%%
\section{Lepton mass spectra}

The mass relations of quarks and leptons (and also
 their superpartners) are generated in our model. 
A main 
 purpose of this paper is to propose a new
 mechanism of generating the bilinear form of the mass spectrum. 
So we briefly show 
 the derivation of the
 charged-lepton mass relation in Eq.(\ref{(1.2)}). 
%However, since one of the motivations for the present
%mechanism was in the VEV relation (1.1), 
% we would like to comment 
% on the case of the charged lepton sector.
%In this section, we does not derive the VEV relation (1.1),
% but, inversely, 

%We 
% investigate what form of $W_\Phi$
% in Eq.(\ref{(2.1)}) is required in order to
% obtain the VEV relation Eq.(\ref{(1.1)}). 

For our goal, 
 we take the simplest 
% the simplest example 
% most straightforward way
% is to assume  the form of
  superpotential $W_\Phi$ 
 as\footnote{
We assume $R$- and $Z_2$-symmetries are
 (spontaneously) broken
 only in $W_\Phi(\Phi_f)$ sector, in which 
 $R(\lambda_d)=R(m_d)=2$ and $Z_2$-parity odd $\lambda_d$ 
 are induced from 
 VEVs of some unknown 
 fields possessing $R$-charges
 and $Z_2$-parity odd.}
\be
W_\Phi(\Phi_d) = \lambda_d {\rm Tr}[\Phi_d] \left( {\rm Tr}[\Phi_d\Phi_d] 
-\frac{2}{9}({\rm Tr}[\Phi_d])^2 \right)
 +m_d {\rm Tr}[\Phi_d\Phi_d] .
\label{(3.1)}
\ee
The form of the cubic term ($\lambda_d$-term) in Eq.(3.1) is equivalent to
\be
\lambda_d \left( {\rm Tr}[\Phi_d\Phi_d\Phi_d] - 
{\rm Tr}[\Phi_d^{(8)}\Phi_d^{(8)}\Phi_d^{(8)}] \right) ,
\label{(3.2)}
\ee
where $\Phi_d^{(8)} = \Phi_d - {\rm Tr}[\Phi_d]/3$.
We assume to
 drop $\Phi^{(8)}_d\Phi^{(8)}_d\Phi^{(8)}_d$-term
 from the $U(3)_F$-invariant cubic term
 ${\rm Tr}[\Phi_d\Phi_d\Phi_d]$. 
It might be possible by imposing
 an additional symmetry as shown in Ref.\cite{Koide0705}.\footnote{
Another $Z_2$-parity
 with $+1$ for $\Phi_d^{(1)}= {\rm Tr}[\Phi_d]/3$ and
 $-1$ for $\Phi_d^{(8)}$ worked well. %has played important role. 
} 
%Thus, we take a standpoint that 
% the form (3.1) is just a phenomenological
% assumption.

{}From Eq.(\ref{(3.1)}), we obtain
\be
\frac{\partial W_\Phi}{\partial \Phi_d} = 0 =
c_1^d \Phi_d + c_0^d {\bf 1} ,
\label{(3.3)}
\ee
where
\bea
c_1^d &=& 2(m_d+ \lambda_d {\rm Tr}[\Phi_d]) ,
\label{(3.4)}\\
c_0^d &=& \lambda_d \left({\rm Tr}[\Phi_d\Phi_d]-\frac{2}{3}
({\rm Tr}[\Phi_d])^2 \right) .
\label{(3.5)}
\eea
The coefficients $c_1^d$ and $c_0^d$ in Eq.(\ref{(3.3)})
 must be zero\cite{previous} in order to 
 obtain 
 non-zero and non-degenerate eigenvalues of 
 $\langle\Phi_d\rangle$. 
Then, the condition $c_0^d=0$ just gives the VEV-relation 
 of Eq.(\ref{(1.1)}) 
in the diagonal basis of $\langle\Phi_d\rangle$. 

We should also notice that 
 Eqs.(\ref{(2.20)}) and (\ref{(2.21)}) require one
 relation  among
 six parameters, $\lambda_A, \lambda_B, \mu_d, \xi_d$,
 and $\lambda_d, m_d$ in $W_\Phi (\Phi_d)$     
 as 
\be
 {\lambda_A\mu^2_d\over \lambda_A^2+\lambda_B^2}
 {\rm Tr}[\xi_d]={2\over 3} {m_d^2\over \lambda_d^2}. 
\label{(2.21a)}
\ee
Taking the same order of $\mu_d$ and $m_d$ might be 
 natural for the setup of the present model.

%%%%%%%%%%%%%%%%%
\section{Slepton mass spectra}

Now let us investigate
 the slepton mass spectra in our model. 
The 
 K\"{a}hler potential of ${\cal O}(\Lambda^{-2})$ 
 derives 
 the SUSY breaking slepton masses
\bea
{\cal K}_L &=& 
\frac{1}{\Lambda^2} 
\left(
{\rm Tr}
\left[\left|
(y_1^d A_d+ y_2^d B_d)L
\right|^2
\right] +
{\rm Tr}
\left[
\left|y_3^d A_d + y_4^d B_d\right|^2
\right]
{\rm Tr}
[\left|L\right|^2]
\right. \nonumber \\
&&+
{\rm Tr}
\left[\left|
(y_1^u A_u+ y_2^u B_u)L
\right|^2
\right] +
{\rm Tr}
\left[
\left|y_3^u A_u + y_4^u B_u\right|^2
\right]
{\rm Tr}
[\left|L\right|^2] \nonumber \\
&&+
{\rm Tr}
\left[\left|
(y_5^d A_d+ y_6^d B_d)E
\right|^2
\right] +
{\rm Tr}
\left[
\left|y_7^d A_d + y_8^d B_d\right|^2
\right]
{\rm Tr}
[\left|E\right|^2] \nonumber \\
&&+
\left.
{\rm Tr}
\left[\left|
(y_5^u A_u+ y_6^d B_u)N
\right|^2
\right] +
{\rm Tr}
\left[
\left|y_7^u A_u + y_8^u B_u\right|^2
\right]
{\rm Tr}
[\left|N\right|^2]
\right) 
\label{(5.1)}
\eea
as well as 
 makes $\langle L \rangle$ and $\langle E \rangle$ 
 zero in the scalar potential. 

Reminding that 
 the charged-lepton masses $M^e_{i}$ have been given by
\be
M_{i}^e = (y_A^d \lambda_B^d-y_B^d \lambda_A^d) 
\frac{\lambda_B^d}{\lambda_A^d} \frac{v_{di}^2}{\Lambda^2}
 \langle H_d \rangle 
\label{(5.2)}
\ee
{}from Eq.(\ref{(2.23)}), 
 the 1st and 2nd terms of Eq.(\ref{(5.1)}) 
 induce the left-handed slepton masses as 
\be
 \left( 
\frac{y_1^d \lambda_B^d -y_2^d \lambda_A^d}
{y_A^d \lambda_B^d -y_B^d \lambda_A^d}
\right)^2 \frac{\Lambda^2}{\langle H_d \rangle^2}\;
\tilde{e}_L{}_i\; (M^e)^2_{i}\; \tilde{e}_L{}_i
+
\left( 
\frac{y_3^d \lambda_B^d -y_4^d \lambda_A^d}
{y_A^d \lambda_B^d -y_B^d \lambda_A^d}
\right)^2 \frac{\Lambda^2}{\langle H_d \rangle^2} \;
\tilde{e}_L{}_i  \tilde{e}_L{}_i \sum_k (M^e)^2_k \; .  
\label{(5.3)}
\ee
The 1st term induces generation-dependent 
 masses (proportional to 
 the charged-lepton squared masses) 
 while the 2nd term gives 
 the universal soft mass. % to all generations. % the sfermion spectrum. 
The neutrino masses $M^{\nu}_i$ are given by the similar
 equation 
 to Eq.(\ref{(5.2)}),  
% with $v_{ui}=\langle (\Phi_u)_{ii}\rangle$. % on the
% diagonal basis of $\langle \Phi_{u}\rangle$.
 and the matrix $\langle \Phi_{u}\rangle$ is
 not diagonal in the diagonal
 basis of $\langle \Phi_{d}\rangle$ 
 because the neutrino mixing matrix $U_\nu$ is not 
 $U_\nu = {\bf 1}$.
So the slepton mass matrix from the 3st and 4th terms in
 Eq.(\ref{(5.1)}) is
 not diagonal in general. 
Here we take a standpoint that neutrinos are Dirac 
 particles with tiny  Dirac masses. 
It is because 
 an introduction of Majorana masses 
 of right-handed neutrinos heavier than 
 $\Lambda$ might be unnatural\footnote{
We might also take another standpoint
 that the origin of Majorana masses is 
 beyond our model and can be heavier than
 $\Lambda$. 
In this case the following results 
 are changed. 
}       
({\it See} Eq.(\ref{(5.6)})).
In this case 
 a small value of $\mu^2_u$ is required 
 which induces 
 the small values of $\langle \Phi_u\rangle^2$ as shown
 in Eq.(\ref{(2.21)}), and then 
 non-degenerate effects from the 
 neutrino sector are negligibly small 
 due to 
 the lepton mass relation 
 $m_\nu^2{}_i/m_{ei}^2 < 10^{-14}$ (even 
 for the inverted-hierarchical neutrino mass spectrum). 
Then   
%Thus, %Anyhow, in this setup,
 the contributions from $A_u$ and $B_u$ in the 3rd and 4th terms  
 in Eq.(\ref{(5.1)}) can be neglected.\footnote{
We do not consider the case of 
  $(y^\nu_i/y^e_i)^2 > 10^{14}$ ($i=1,2,3$) 
 and accidental cancellations among $y$'s, $\lambda$'s, 
 and $\langle\Phi\rangle$'s. 
}

Therefore the left-handed 
 slepton masses $(\tilde{m}_{LL})_i^2$ have 
 the form\footnote{
Here 
$k_L =  [({y_1^d \lambda_B^d -y_2^d \lambda_A^d})/
({y_A^d \lambda_B^d -y_B^d \lambda_A^d})]^2 
({\Lambda}/\langle H_d\rangle)^2$, and so on.  
} 
\be
 (\tilde{m}_{LL})_i^2 = k_L [(M^e)_{i}^2+m_{L0}^2] 
 %+{\cal O}(k_L' m_\nu^2{}_i),
 \label{(5.4)}
\ee
where 
% and $k_L' = {\cal O}(\Lambda^2/\langle H_u\rangle^2)$.
$m_{L0}^2$ is the universal soft mass for all three
 generations.  
It should be noticed that 
 non-degenerate masses can dominate 
 the universal masses, 
 in which charged-slepton masses 
 are almost proportional to
 the charged-lepton masses. 
%The two terms in Eq.(\ref{(5.4)}) are 
% comparable for 
% $y_{1,2}^d \sim y_{3,4}^d$.  
The right-handed charged-slepton 
 masses are 
 also calculated  from Eq.(\ref{(5.1)}) as 
%The result is also presented by the form
\be
 (\tilde{m}_{RR})_i^2 = k_R [(M^e)_i^2+m_{R0}^2] .
 \label{(5.5)}
\ee
Notice that the left-right mixing 
 masses $\tilde{m}_{LR}^2$ %are zero
%, since  the term 
% ${\rm Tr}\left[ A_d L H_d E \right]_F$ and 
% ${\rm Tr}\left[ B_d L H_d E \right]_F$
 are zero due to the (approximate) 
 $R$-symmetry. %charge assignment. 
As for sneutrino sector, 
 right-handed sneutrinos have masses of 
 ${\cal O}(M^\nu\Lambda/\langle H_u \rangle)$, 
% which tend to be the NLSP. 
 for example, right-handed $\tau$-sneutrino 
 has a mass of ${\cal O}(100)$ eV in case of
 $\Lambda \sim 10^5$ GeV. 
Due to the tiny 
 neutrino Yukawa couplings, 
 the FCNC processes induced by
 the (s)neutrinos are suppressed.  
Anyhow %an existence of 
 the light sneutrinos
% is one of the predictions of this model, and it 
 might be interesting 
 for the cosmology. 
%universal masses are dominant due to the 
% smallness of Dirac masses as shown above. 

We emphasize again that 
 the charged-lepton and charged-slepton 
% and 
% left-handed sneutrino
 mass matrices
 are diagonalized simultaneously in the flavor space.
% the slepton mass matrix becomes 
% diagonal on the basis where the charged lepton mass 
% matrix is diagonal. 
Therefore, 
 the FCNC processes 
 in the lepton sector, for example  
 $\mu \rightarrow e\gamma$, are suppressed, 
 although 
 the charged-sleptons have non-degenerate masses in general.  
%In the specific case, 
% slepton masses 
% can be proportional to 
% the charged-lepton squared
%  masses. 
%As shown in Eqs.(\ref{(5.4)}) and (\ref{(5.5)}), 
% the charged-slepton mass spectrum 
% is characteristic, that is, 
% the mass squared is propotinal to
% (charged-lepton mass)$^2$+ (an universal mass)$^2$. 
%The charged-lepton mass relation Eq.(\ref{(1.2)}) 
% exists 
% in (charged-lepton  mass)$^2$. 
%Notice that  
% the sleptons are not necessarily degenerate, 
% and diagonalized simultaneously with 
% the lepton sector. 

As for the gaugino masses,
 it is difficult to generate the suitable scale
 of them. 
It is because the $R$-symmetry is broken only in 
 $W_\Phi$ sector. 
(Reminding 
 that gaugino masses require both SUSY and
 $R$-symmetry breakings, only higher order 
 operators can induce
 gaugino masses in the present
 model.) 
So here we assume that the gaugino masses are 
 induced another source, such as 
 moduli $F$-terms. 
The $\mu$-term is also assumed to be
 induced from another mechanism.

\vspace{3mm}

Here let us fix the scale of
 $\Lambda$ in our model. 
We take the 
 soft SUSY breaking masses as 
\be
F/\Lambda \sim {\cal O} (1) \hbox{ TeV}
\label{(4.6)}
\ee
 ($\tilde{m}_\tau\simeq 1$ TeV),     
 while %On the other hand 
 the $\tau$-Yukawa should be 
\be
F/\Lambda^2\sim{\cal O}(10^{-2}).  
\label{(4.7)}
\ee
Combining Eqs.(\ref{(4.6)}) and (\ref{(4.7)}), 
 the scale of $\Lambda$ should 
 be  
\be
\Lambda \sim {\cal O}(10^5) \, {\rm GeV} . 
\label{(5.6)}
\ee
%{}from Eq.(\ref{(5.3)}). 
%Thus we can not regard $M$ as the 4 dimensional 
% Planck mass $M_{Pl}$.
One example of derivation $\Lambda$ 
 is considering %coming from 
 the large extra dimensional  
 theory\cite{ArkaniHamed:1998rs}, 
 in which the $\Lambda$ is regarded as 
 the $D$-dimensional Planck
 scale $M_*$ with a relation of         
\be
%\frac{1}{8\pi}
M_P^2 = M_*^{2+d}(2\pi R)^d \; ,
\label{(5.7)}
\ee
% is regarded as $\Lambda$. 
where $d$ is a 
 number of the extra dimension 
 ($D=4 +d$). %, and 
% $M_P$ is the four-dimensional 
% Planck scale. 
$M_*\sim 10^5$ GeV means 
 $R\sim 10^8\; {\rm GeV}^{-1} (\sim 10^{-7}\, {\rm m})$ 
 in the case of  $D=6$. 
%This $R$ is experimentally acceptable.  
(The $D=5$ case is experimentally excluded.)
The present model 
 requires a setup in which all fields except for 
 gravity multiplets are localized on
 the 4-dimensional brane.  

\vspace{3mm}

In  
  a short summary 
 we present the mass spectra of the model.  
The  input parameters are
  $m_\tau \sim 1$ GeV, 
 $m_{\tilde{\tau}} \sim 10^3$ GeV 
 and $m_{\nu 3} \sim 10^{-10}$ GeV. 
Then the outputs are 
 $\langle\Phi\rangle_d \sim 10^4$ GeV, 
 $\langle\Phi\rangle_u \sim 10^{-1}$ GeV  
 and $\Lambda\sim 10^5$ GeV. 
The following table shows 
 the order of mass spectra of 
 $U(3)_F$-nonet fields. 
\begin{center}
\begin{tabular}{|c|c|c|c|}\hline
 fields & VEV & fermion masses & scalar masses \\ 
\hline
$\Phi_f$ & $\langle\Phi\rangle_f $ & $\langle\Phi\rangle_f $
& $\langle\Phi_f\rangle $ \\
$C_f$   & 0 & $\langle\Phi\rangle_f $ & $\langle\Phi_f\rangle $ \\
$C'_f$    & 0 & 0 & $\langle\Phi_f\rangle^2/\Lambda$ \\ 
%\tau    & 0 & 1 GeV & $10^3$ GeV \\ 
%
\hline
\end{tabular}
\end{center}
We should notice that the gravitino is the next-LSP (NLSP), 
 $m_{3/2} \sim 0.1$ eV, and the mass of 
 right-handed $\tau$-sneutrino is 
 of order 100 eV.
The lightest right-handed sneutrino is the
 next-to-next-LSP (NNLSP) in this model.\footnote{
The cosmological studies might be interesting. 
}

\vspace{3mm}

Finally we comment on 
 the case of 
 introducing a soft term 
\be
V_{\rm soft}\sim m_{3/2}\: \mu_f^2 \: {\rm Tr}[\xi_f A_f] +{\rm h.c.} ,
\ee
in the supergravity (SUGRA) setup. 
% and $m_{3/2} (\sim F/M_P)$ is the gravitino
% mass.  
Notice that this term breaks 
 the $R$-symmetry explicitly, and then  
 the vacuum is shift to 
 non-vanishing $\langle A \rangle$ and
  $\langle B \rangle$ 
 $\sim \Lambda^2/M_P$. 
Then 
 massless eight 
 $\psi_{C'}$ gain 
 ${\cal O}(m_{3/2})$ masses 
 through the
 non-canonical K\"{a}hler ${\cal K}_1$ in Eq.(\ref{(2.4)}).  
In this case the processes 
 which have $\psi_{C'}$ in the external 
 lines are no longer vanished. 
%NLSP is no longer the stable
% due to nonvanshing mass of $\psi_{C'}$. 
For example,
 a slepton decays to 
 a lepton and $\psi_{C'}$, such 
 as  
 $L \rightarrow \psi_{C'_d} +\psi_E$ 
 through the K\"{a}hler interaction in Eq.(\ref{(4.3)}). 
% $(\chi^0)$ decays
% to $\psi_{C'}$
% through 
% a process of 
% $\tilde{\chi}^0 \rightarrow {\psi_L} +L \rightarrow 
% \psi_L+(\psi_{C'_d} +\psi_E)$. 
The partial decay width of this process is 
 roughly estimated\footnote{
More accurate estimations should be needed 
 to check whether this process disturbs 
 the big bang necleon synthesis or not.
} as 
 $\Gamma\geq m_{3/2}^2\langle H_d \rangle^2
 \tilde{m}_E/\Lambda^4$.
% this decay process can finish 
% before 1 sec in the early
% universe. 
% does not destroy the big bang necleon
% synthesis. 
%Whether $\psi_{C'}$ can be the dark matter
% or not needs further estimations\cite{}. 
%We need another dark matter candidate. 
%Due to non-zero $\langle A \rangle$ and
%  $\langle B \rangle$.   
Although the gaugino masses can be 
 generated by the gauge mediation
 with the additional
 messenger fields\cite{GM}, 
 they are too light and still need
 another SUSY breaking source. 

% scenario\cite{}
% with the additional messenger fields\cite{}. 

%LPS is not Sperticle.
%We need another dark matter candidate. 
%The decay is faster than 
% 1 sec when $\Lambda \sim 10^5$ GeV. 

%$V_[soft]$ case change the situation.

%The decay is before 
% BBN 1sec. 

%%%%%%%%%%%%%%%%%

\section{Summary and discussions}

We have investigated new mechanism which
 induces flavor mass spectrum by the 
 $F$-components of $U(3)_F$-nonet superfields. 
%The model is based on an O'Raifeartaigh-type 
% SUSY breaking model. 
Fermion masses
 are generated through the K\"{a}hler potential,
 and then the desired bilinear form of the mass spectrum
 is realized as 
 $(M^e)_{ij} \propto \sum_k \langle\Phi_{dik}\rangle
 \langle\Phi_{dkj}\rangle$.
This can induce the 
 interesting 
 charged-lepton mass relation 
 Eq.(\ref{(1.2)}) by using  a particular 
 form of $W_\Phi(\Phi_d)$. 

In the nonet superfields,  
 eight fermions of 
 $(\psi_{C'})$ %($f=u,d$)
 remain as 
 massless particles. 
However 
 smallness of their couplings and 
 no chiral-flip of them in the loop diagrams 
 suppress the FCNC processes. 
The interactions
 with $\psi_{C'}$ in the external lines 
 vanish so that the decay processes to $\psi_{C'}$ 
 are strongly suppressed.

The charged-slepton mass spectrum 
 is non-degenerate in general and can be even hierarchical
 (proportional to the charged-lepton
  masses in the specific case).
The FCNCs in the lepton sector 
 are suppressed  
 since 
 the {charged-lepton} and {slepton} (except for 
 right-handed sneutrino) 
 mass matrices 
 are diagonalized simultaneously in the flavor space.
The right-handed {sneutrino}s are light with the similar 
 ratio to 
 the {lepton} sector
 ($\tilde{m}_{\nu_R}$/$\tilde{m}_e \sim m_\nu$/$m_e$). 
Due to the tiny 
 neutrino effective Yukawa couplings, 
 the FCNC processes induced by
 the (s)neutrinos are suppressed.  
The gravitino mass is about 
 0.1 eV 
 in case of
 $\Lambda \sim 10^5$ GeV, and  
 the setup of our model fits 
 the large extra dimension scenario. 
We have also comment on the present 
 model in the SUGRA setup, in which  
 massless $\psi_{C'}$
 obtain masses of ${\cal O}(m_{3/2})$.

%%%%%%%%%%%%%%%%%%%%%%%%%%%%%%%%%%%%%%%%%%%%%
\vspace{1cm}
%\newpage

\centerline{\large\bf Acknowledgments} 

\vspace{5mm}
The authors thank Y.~Hyakutake, T.~Yamashita, 
 K.~Oda, Y.~ Sakamura, M.~ Tanaka, 
 N.~Okada, K.~Yoshioka, M.~Tanimoto, and S.~Matsumoto 
 for helpful discussions.
 One of the authors (YK) is also grateful to J.~Sato, M.~Yamanaka
 and T.~Nomura for valuable comments and their kind hospitalities
 at Saitama University.  
 N.H. 
 and Y.K.
 are supported by the Grant-in-Aid for
 Scientific Research, Ministry of Education, Science and 
 Culture, Japan (No.16540258, No.17740146) and
 (No.18540284), respectively.

%%%%%%%%%%%%%%%%%%%%%%%%%%%%%%%%%%%%%%%%%%%%

%%%%%%%%%%%%%%%%%%%%%%%%%%%%%%%%%%%%%%%%%%%%


\vspace{.5cm}
\begin{thebibliography}{99}
%

\bibitem{Koide90} Y.~Koide, Mod.~Phys.~Lett. {\bf A5} (1990) 2319.

 \bibitem{Koidemass} Y.~Koide, Lett.~Nuovo Cimento {\bf 34} (1982)  201;
Phys.~Lett. {\bf B120} (1983) 161;
Phys.~Rev. {\bf D28} (1983) 252.
%

%
\bibitem{PDG06} Particle Data Group, J.~Phys.~G {\bf 33} (2006) 1.
%
\bibitem{FCNC}
S.~Glashow and S.~Weinberg, Phys.~Rev. {\bf D15} (1977) 1958.
%
%\bibitem{S3} S.~Pakvasa and H.~Sugawara, Phys.~Lett. {\bf B73}
%(1978) 61; 
%H.~Harari, H.~Haut and J.~Weyers, Phys.~Lett. {\bf B78} (1978) 459; 
%J.-E.~Fr\`{e}re, Phys.~Lett. {\bf B80} (1978) 369; 
%E.~Derman, Phys.~Rev. {\bf D19} (1979) 317; 
%D.~Wyler, Phys.~Rev. {\bf D19} (1979) 330.
%
%\bibitem{UnivSeesaw} 
%Z.~G.~Berezhiani, Phys.~Lett.~{\bf 129B} (1983) 99;
%Phys.~Lett.~{\bf 150B} (1985) 177;
%D.~Chang and R.~N.~Mohapatra, Phys.~Rev.~Lett.~{\bf 58} (1987) 1600; 
%A.~Davidson and K.~C.~Wali, Phys.~Rev.~Lett.~{\bf 59} (1987) 393;
%S.~Rajpoot, Mod.~Phys.~Lett. {\bf A2} (1987) 307; 
%Phys.~Lett.~{\bf 191B}, 122 (1987); Phys.~Rev.~{\bf D36} (1987) 1479;
%K.~S.~Babu and R.~N.~Mohapatra, Phys.~Rev.~Lett.~{\bf 62} (1989) 1079; 
%Phys.~Rev. {\bf D41} (1990) 1286; 
%S.~Ranfone, Phys.~Rev.~{\bf D42} (1990) 3819; 
%A.~Davidson, S.~Ranfone and K.~C.~Wali, 
%Phys.~Rev.~{\bf D41} (1990) 208; 
%I.~Sogami and T.~Shinohara, Prog.~Theor.~Phys.~{\bf 86} (1991) 1031;
%Phys.~Rev. {\bf D47} (1993) 2905; 
%Z.~G.~Berezhiani and R.~Rattazzi, Phys.~Lett.~{\bf B279} (1992) 124;
%P.~Cho, Phys.~Rev. {\bf D48} (1993) 5331; 
%A.~Davidson, L.~Michel, M.~L.~Sage and  K.~C.~Wali, 
%Phys.~Rev.~{\bf D49} (1994) 1378; 
%W.~A.~Ponce, A.~Zepeda and R.~G.~Lozano, 
%Phys.~Rev.~{\bf D49} (1994) 4954.
%
\bibitem{Froggatt} C.~Froggatt and H.~B.~Nielsen, Nucl.~Phys. {\bf B147}
  (1979) 277.
%
\bibitem{Koide0705}
Y.~Koide, JHEP {\bf 08} (2007) 086. % arXiv hep-ph/0705.2275 (2007) 
%
\bibitem{Ma-PLB07}
 E.~Ma, Phys.~Lett.~{\bf B649} (2007) 287.
%
\bibitem{Murayama}
N.~Arkani-Hamed, L.~Hall, H.~Murayama, S.~Smith and N.~Weiner,
Phys.~Rev. {\bf D64} (2001) 115011.
%
\bibitem{ORft}
L.~O'Raifeartaigh, Nucl.~Phys. {\bf B96} (1975) 331.
%
%\bibitem{solar} 
%B.~Aharmim {\it et al.} SNO Collaboration,
%Phys.~Rev. {\bf C72} (2005) 055502.
%T.~Araki  {\it et al.} KamLAND Collaboration, 
%Phys.~Rev.~Lett. {\bf 94} (2005) 081801.
%
%\bibitem{atm} 
%D.~G.~Michael {\it et al.}  MINOS Collaboration, Phys.~Rev.~Lett.
%{\bf 97} (2006) 191801.
%Also see, J.~Hosaka {\it et al.} the Super-Kamiokande Collaboration,
%arXiv: hep-ex/0604011.
%
%\bibitem{seesaw} 
%P. Minkowski, Phys.~Lett. {\bf 67B} (1977) 421.
%

\bibitem{previous}
  N.~Haba and Y.~Koide,
  %``New Origin of a Bilinear Mass Matrix Form,''
  Phys.\ Lett.\  B {\bf 659} (2008) 260. 



\bibitem{ArkaniHamed:1998rs}
  N.~Arkani-Hamed, S.~Dimopoulos and G.~R.~Dvali,
  %``The hierarchy problem and new dimensions at a millimeter,''
  Phys.\ Lett.\  B {\bf 429} (1998) 263.

\bibitem{GM}
M. Dine and A. E. Nelson, 
%gDynamical supersymmetry breaking at low-energies,h 
Phys.\ Rev.\ D {\bf 48} (1993) 1277.  
%
%%\bibitem{Brannen}
%C.~Brannen, 
%http://brannenworks.com/MASSES2.pdf, (2006).
%
%\bibitem{Mohapatra-S4} C.~Hagedorn, M.~Linder and R.~N.~Mohapatra,
%JHEP, {\bf 0606} (2006) 042. %[hep-ph/0602244].
%
%\bibitem{tribi} 
% L.~Wolfenstein,
  %``Oscillations Among Three Neutrino Types And CP Violation,''
%Phys.~Rev. {\bf D18} (1978) 958;
%S.~Pakvasa and H.~Sugawara, Phys.~Lett. {\bf B82} (1979) 105;
%Y.~Yamanaka, H.~Sugawara and S.~Pakvasa, Phys.~Rev. {\bf D25} (1982) 1895;
%P.~F.~Harrison, D.~H.~Perkins and W.~G.~Scott,
% Phys.~Lett. {\bf B458} (1999) 79;
% Phys.~Lett. {\bf B530} (2002) 167;
%R.~N.~Mohapatra and S.~Nussinov,
% Phys.~Rev.  {\bf D60} (1999) 013002;  
%Z.~Z.~Xing, Phys.~Lett. {\bf B533} (2002) 85;
%P.~F.~Harrison and W.~G.~Scott,
% Phys.~Lett. {\bf B535} (2003) 163;
%Phys.~Lett. {\bf B557} (2003) 76;
%E.~Ma, Phys.~Rev.~Lett. {\bf 90} (2003) 221802;
%C.~I.~Low and R.~R.~Volkas, Phys.~Rev. {\bf D68} (2003) 033007;
%X.-G.~He and A.~Zee,  Phys.~Lett. {\bf B560} (2003) 87;
%R.~N.~Mohapatra, S.~Nasri and H.~B.~Yu, Phys.~Lett. {\bf B639} (2006) 318;
%N.~Haba, A.~Watanabe and K.~Yoshioka,
 %``Twisted flavors and tri/bi-maximal neutrino mixing,''
% Phys.~Rev.~Lett. {\bf 97} (2006) 041601.
%
%\bibitem{KoideJPG07}
%Y.~Koide, J.~Phys. {\bf G34} (2007) 1653. 
 
\end{thebibliography}
\end{document}